# Realization of Phonon FETs in 2D material through Engineered Acoustic Mismatch

H. F. Feng, Z. Y. Xu, B. Liu, and Zhi-Xin Guo*

*State Key Laboratory for Mechanical Behavior of Materials, School of Materials Science and Engineering, Xi'an Jiaotong University, Xi'an, Shanxi, 710049, China.*

*zxguo08@xjtu.edu.cn

**Abstract**

Field-effect transistors (FETs) predominantly utilize electrons for signal processing in modern electronics. In contrast, phonon-based field-effect transistors (PFETs)—which employ phonons for active thermal management—remain markedly underdeveloped, with effectively reversible thermal conductivity modulation posing a significant challenge. Herein, we propose a novel PFET architecture enabling reversible thermal conductivity modulation. This design integrates a substrate in the central region with a two-dimensional (2D) material to form an engineered junction, exploiting differences in out-of-plane acoustic phonon properties to regulate heat flow. Molecular dynamics simulations of a graphene (Gr)/hexagonal boron nitride (h-BN) junction demonstrate a substantial thermal conductivity reduction—up to 44-fold at 100 K. The effect is maintained at room temperature and across diverse substrates, confirming robustness. This work establishes a new strategy for dynamic thermal management in electronics.

# Introduction

Electron-based field-effect transistors (FETs) underpin modern electronics, serving as fundamental components for signal processing [1-6]. In contrast, their phononic counterparts—phonon-based field-effect transistors (PFETs) designed for active thermal management—remain markedly underdeveloped. A central challenge lies in achieving reversible thermal conductivity modulation. Despite decades of research into controlling thermal transport in solids, most existing methods are irreversible [7-9]; material thermal conductivity becomes fixed post-synthesis, severely limiting dynamic control [10-13]. Thus, developing technologies capable of actively and reversibly controlling thermal conductivity is crucial, especially for devices requiring dynamic thermal management [14-18].

In recent years, reversible modulation of thermal conductivity has sparked considerable research interest [19-26]. Studies have shown that external fields can induce thermal switching via phase transition or interface reconstruction in ferroic (ferroelectric/ferromagnetic) materials [27-33]. However, strategies for dynamic thermal control without inducing such structural transformations remain elusive. Our prior work established that substrate-mediated mechanical perturbation significantly modulates both in-plane and out-of-plane phonon contributions to thermal transport in two-dimensional (2D) materials [34]. Crucially, enhanced interlayer coupling in a 2D material induces frequency shifts in out-of-plane acoustic (ZA) phonon modes, altering thermal conductivity. While this method has a limited impact on the regulation of thermal conductivity, the significant regulation of interlayer coupling on phonon modes provides a new avenue for thermal conductivity control.

In this paper, we propose a novel PFET architecture based on engineered 2D materials. By integrating a substrate locally with a 2D material, we create spatially distinct regions with varying out-of-plane acoustic phonon properties. This design enables effective control of thermal conductivity by modulating the phonon modes mismatch between distinct spatial regions, realizing phonon field-effect modulation. To validate this concept, we employ molecular dynamics simulations to investigate the effect of a hexagonal boron nitride (h-BN) substrate junction on the thermal conductivity of graphene (Gr). By adjusting the interlayer distance, we can modify the interlayer coupling strength, achieving sustained modulation of thermal conductivity. It is noted that at 100 K, we observe a 44-fold reduction in thermal conductivity. The method also proves effective at room temperature and across different substrates, demonstrating its versatility and reliability. The proposed PFET combines three key advantages: miniaturization, high-frequency switching capabilities, and operational longevity, which shows a novel direction in thermal management research.

## Results and discussions

Our previous studies have shown that introducing substrates on both sides of 2D materials, combined with enhanced interlayer coupling, significantly alters the phonon spectrum, particularly for out-of-plane phonon modes [34]. For example, in a monolayer Gr completely encapsulated by h-BN substrates (see Fig. S1 in the Supplemental Material [35]), the phonon spectrum for in-plane modes remains largely unaffected with the introduction of h-BN [Fig. 1(a)]. Nonetheless, the ZA mode exhibits a significant frequency upshift due to increased coupling with the surrounding h-BN layers. This shift scales positively with interfacial coupling strength and inversely with $d_{int}$, in agreement with our previous findings. To better understand these phonon changes, we calculate the phonon density of states (DOS) under varying interlayer coupling conditions, separately analyzing in-plane and out-of-plane contributions. As shown in Fig. 1(b), the phonon DOS of in-plane modes remains nearly constant with the introduction of h-BN. In contrast, the phonon DOS of out-of-plane modes develops a low-frequency band gap due to the upshift of the ZA mode [Fig. 1(c)]. This gap widens and shifts to higher frequencies as interlayer coupling increases, diverging significantly from the behavior of freestanding Gr. Given that the low-frequency ZA mode plays a dominant role in heat transport within 2D materials, these changes are expected to have a substantial impact on thermal conductivity. This highlights interlayer coupling as an effective and tunable mechanism for controlling phonon transport, offering a promising strategy for modulating thermal conductivity in layered nanomaterials.

Based on these differences in phonon properties with varying interlayer couplings, we further propose a series-connected structure for a PFET to modulate thermal conductivity. As illustrated in Fig. 1(d), the device consists of a heat-conducting layer (Gr) with a central region covered by h-BN, forming a junction between the covered and uncovered Gr. By adjusting the interlayer distance at the junction, one can effectively control the coupling strength in this region, thereby modifying the phonon DOS at the junction. This creates a significant mismatch in DOS between the junction and the uncovered Gr, selectively impeding the transmission of low-frequency phonons, primarily from ZA modes. Hence, h-BN in the PFET functions as a phononic gating, analogous to the role of a gate dielectric in the electronic FET [1-3]. Since the low-frequency ZA mode is major contributor to thermal conductivity in 2D materials [47, 48], this PEET configuration enables effective control over thermal conductivity via changing the interlayer coupling strength ($d_{int}$).

To validate the above mechanism, we conduct molecular dynamics (MD) simulations on the thermal conductivity of PFET. Our calculations employ both homogeneous nonequilibrium molecular dynamics (HNEMD) and non-equilibrium molecular dynamics (NEMD) methods [49,50]. Given the strong dependence of low-frequency phonons on interlayer coupling and their dominant role at low temperatures, we carry out simulations at 100 K. Figure 2(a) presents the temperature distribution under varying interlayer couplings obtained from the NEMD simulation. A general feature is that introducing the junction leads to a sharp temperature jump at the junction, the magnitude of which increases with the junction spacing $d_{int}$ decreasing. This temperature jump reflects a sharp rise in thermal resistance, indicating that the presence of the junction effectively hinders the thermal transport, and the stronger interlayer coupling leads to a more pronounced thermal conductivity decrease in the PFET.

Then, we systematically investigate the thermal conductivity of PEET under different interlayer coupling strengths using the MD simulations. To accurately resolve the different directional mode contributions to heat transport, we employ the HNEMD method. This approach allows us to separate the thermal conductivity contributions from in-plane ($\kappa_{in}$) and out-of-plane phonons ($\kappa_{out}$) and effectively incorporates quantum effects at low temperatures. In calculating thermal conductivity, it is essential to define the effective thickness. For the junction structures, we take the interlayer distance ($d_{int}$)—the distance between the gating and heat-conducting layer—as the effective thickness. Specifically, the equilibrium distance between Gr and h-BN is set to 3.4 Å, which represents Van der Waals interactions. To ensure consistency across all configurations, a standard thickness of 3.4 Å is applied in our calculations.

As shown in Fig. 2(b), introducing a substrate and increasing the interlayer coupling leads to a significant and continuous reduction in total thermal conductivity—from 2889.2 W/mK in freestanding Gr (in the absence of interlayer interactions) to 406.1, 177.0, and 65.7 W/mK in PFETs with $d_{int}$ values of 3.4, 3.1, and 2.8 Å, respectively. Conversely, reducing the coupling strength by increasing $d_{int}$ to 3.5 and 3.6 Å results in thermal conductivities of 662.5 and 2174.8 W/mK (Fig. S2), respectively. These results demonstrate that weaker interlayer coupling leads to higher thermal conductivity, gradually approaching the value of freestanding Gr. Therefore, freestanding Gr without interlayer coupling can be regarded as the "on" state of a thermal switch, while the PFETs with smaller $d_{int}$ values represent the "off" states. The corresponding thermal on–off ratios of 7.1, 16.3, and 44.0 for $d_{int}$ values of 3.4, 3.1, and 2.8 Å, respectively, highlight the strong influence of interlayer coupling on thermal transport. Notably, unlike conventional FETs that offer binary (on/off) switching, this structure allows for multiple, tunable thermal resistance states by simply adjusting the interlayer distance.

Based on the different sensitivities of in-plane and out-of-plane phonons to interlayer interactions, we further separate their contributions to thermal conductivity. As shown in Fig. 2(b), $\kappa_{out}$ in Gr is 2508.9 W/mK, accounting for 86.8% of the total thermal conductivity. However, in the PFET, the presence of interlayer interactions significantly suppresses this contribution. As $d_{int}$ decreases to 3.4, 3.1, and 2.8 Å, the $\kappa_{out}$ drops sharply to 248.2, 58.4, and just 0.6 W/mK, respectively. Particularly, at 2.8 Å, the out-of-plane phonons contribute only 0.9% to the total thermal conductivity, indicating that interlayer interactions nearly eliminate out-of-plane phonon transport. This dramatic suppression mirrors the "off" state in electronic transistors, where conduction is effectively blocked. Notably, the reduction in $\kappa_{out}$ accounts for 88.8% of the total decrease in thermal conductivity, confirming its dominant role in thermal modulation within the PFET. These findings align with the observed changes in the phonon DOS mismatch between the junction and uncovered Gr, further validating that the modulation of out-of-plane phonons at the junction is primarily responsible for the PFET's thermal on–off behavior.

To further explore this mechanism, we calculate the phonon transmission spectra for different interlayer couplings in the junction. As shown in Fig. S3, the transmission spectra of in-plane phonons show minimal variation with changes in interlayer interaction ($d_{int}$), consistent with the characteristics of phonon DOS and thermal conductivity shown in Fig. 1(b) and Fig. 2(b). Nonetheless, for out-of-plane phonons, the transmission spectra exhibit significant differences with varying interlayer interactions. As shown in Fig. 3(a), for out-of-plane phonons in uncovered Gr, transmission spans the entire 0-40 THz frequency range. After introducing interlayer coupling, the transmission in the low-frequency region is nearly blocked, forming a phonon band gap of 0-5 to 0-13 THz. This band gap, which is positively correlated with interface coupling strength, prevents phonons in this frequency range from passing through the junction, thereby reducing their contribution to heat transport. As the junction spacing decreases, the band gap widens, and the transmission near the gap weakens, visually demonstrating the junction's ability to filter low-frequency phonons and impact thermal conductivity. Since low-frequency out-of-plane modes are the primary contributors to thermal conductivity in uncovered Gr [47,48], this result clearly confirms that modulation of out-of-plane phonons through interlayer interactions is the main driver of the thermal on–off ratio in the PFET.

To quantify the impact of the phonon band gap on thermal conductivity, we additionally perform a frequency decomposition of $\kappa_{out}$. As shown in Fig. 3(b), in freestanding Gr, the thermal conductivity contribution decreases with increasing frequency, with low-frequency phonons dominating the $\kappa_{out}$. It is noted that the introduction of the junction creates a phonon band gap in the low-frequency region, leading to a significant reduction in thermal conductivity. Moreover, as the junction spacing decreases, the band gap widens, further suppressing phonon heat transport at low frequencies and causing a marked decrease in total thermal conductivity. This behavior is consistent with the transmission spectra in Fig. 3(a). Notably, at a junction spacing of 2.8 Å, phonons in the 0-20 THz range are almost entirely blocked, demonstrating the junction's effectiveness in modulating thermal conductivity when out-of-plane phonons predominantly govern heat transport.

To evaluate the general applicability of our model, we extend our simulations to additional conditions. First, we examine thermal transport of above the PFET system at room temperature (300 K), as shown in Fig. 4(a). With the addition of interlayer interactions from h-BN in the central region, the total thermal conductivity decreases from 1895.7 W/mK in Gr to 833.1, 511.0, and 281.1 W/mK in PFETs—achieving an on-off ratio of 2.3, 3.7, and 6.7 for $d_{int}$ = 3.4, 3.1, and 2.8 Å, respectively. Although these ratios are smaller than those at 100 K, they are still significant and surpass much of what is reported in previous studies [19,26]. This confirms a consistent, monotonic decline in thermal conductivity with increasing coupling strength, even at higher temperatures, highlighting the broader practical relevance of our approach.

We also evaluate $\kappa_{in}$ to the change of total thermal conductivity in the PFETs at 300 K, as shown in Fig. 4(a). It is found that when interlayer coupling is relatively weak ($d_{int}$ = 3.4 Å), $\kappa_{in}$ shows minimal change. Even at stronger interlayer coupling ($d_{int}$ = 2.8 Å), $\kappa_{in}$ only decreases by 40%. This aligns with relatively stable DOS of in-plane phonons observed across different coupling strengths in Fig. 1(b). In contrast, $\kappa_{out}$ decreases sharply from 1495.5 W/mK in Gr to 40.5 W/mK in PFET with $d_{int}$ = 2.8 Å, representing just 2.7% of its original value. This dramatic reduction marks the "off" state in the FET. Overall, the significant decrease in thermal conductivity is primarily driven by the suppression of out-of-plane phonons, consistent with the observations at 100 K.

To further test the robustness of this modulation strategy, we replace the h-BN gating with Gr [see Fig. S4(a)] and perform the same simulation to calculate the thermal conductivity. As shown in Fig. 4(b), the results closely align with those obtained using h-BN gating, indicating that the gating material has little influence on the overall modulation effect. This suggests that the ability to regulate heat transport is not strongly dependent on the specific choice of substrate. We also examine the effect of changing the heat-conducting layer itself. When h-BN is used as the transport layer [Fig. S4(b)], we again observe a clear and continuous decrease in total thermal conductivity with increasing interlayer coupling [Fig. 4(c)]. The conductivity decreases from 455.6 W/mK in h-BN to 218.1, 164.0, and 134.2 W/mK in PFETs with $d_{int}$ = 3.4, 3.1, and 2.8 Å, respectively, corresponding to on-off ratio of 2.1, 2.8, and 3.4. Although the thermal conductivity modulation effect is less pronounced than in Gr, $\kappa_{out}$ still shows a significant decline, dropping to 4.0% of its original value at $d_{int}$ = 2.8 Å. These findings demonstrate that the proposed approach is broadly applicable across different 2D material systems and is especially effective in cases where out-of-plane phonons dominate thermal transport.

The above results demonstrate that the junction structure effectively modulates the thermal conductivity of monolayer materials. To assess whether this effect extends to multilayer 2D materials, we choose bilayer graphene (BG) as a heat-conducting layer for the validation [see Fig. S4(c) for the device model]. As shown in Fig. 4(d), thermal conductivity also remarkably decreases from 1312.2 W/mK in BG to 788.4.1, 534.0, and 331.6 W/mK in PFETs with $d_{int}$ = 3.4, 3.1, and 2.8 Å, respectively. These correspond to on-off ratio of 1.7, 2.5, and 4.0. Particularly, with a strong interlayer interaction ($d_{int}$ = 2.8 Å), the $\kappa_{out}$ also drops to 6% of its original value, confirming that strong interlayer coupling remains an effective mechanism for suppressing thermal transport in multilayer systems. Together, these results confirm the robustness and versatility of the proposed PFET design for thermal conductivity modulation in both monolayer and multilayer 2D materials.

## Conclusion

This study presents a novel method for dynamically modulating thermal conductivity in 2D materials through the integration of a PFET. By manipulating the interlayer coupling in a junction structure, we demonstrate significant control over phonon transmission, particularly in the out-of-plane direction. Molecular dynamics simulations reveal that, at low temperatures, the thermal conductivity of Gr could be reduced by up to 44 times, with similar results observed at room temperature, confirming the practicality of this approach. Furthermore, this technique is applicable to a range of substrates and material systems, including multilayer 2D materials, extending its potential for a wide variety of applications. The results offer a promising pathway for advancing thermal management in electronic devices, where dynamic and reversible control of thermal conductivity is crucial.

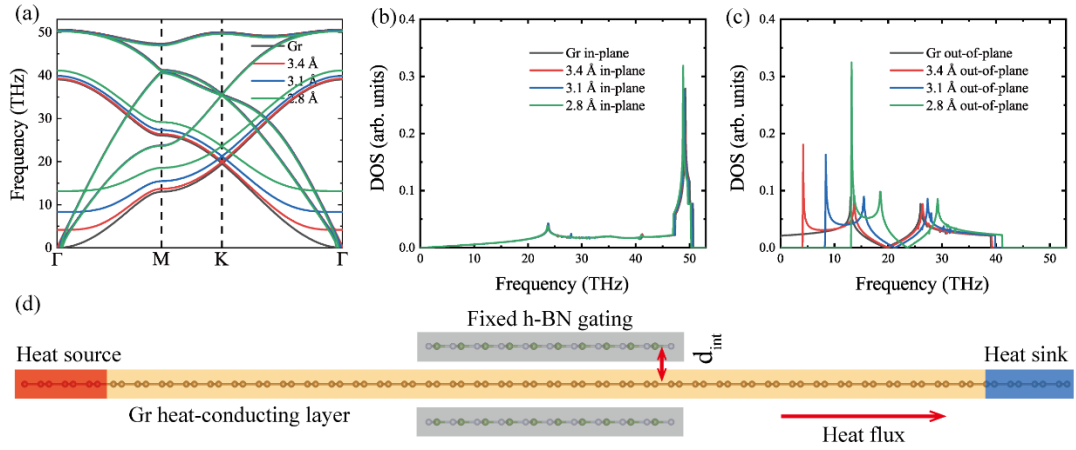

FIG. 1. (a) Phonon spectrum of freestanding Gr and Gr in sandwich structures with different $d_{int}$. (b)-(c): DOS of freestanding Gr and Gr in sandwich structures with different $d_{int}$ (= 3.4 Å, 3.1 Å, 2.8 Å) for (b) in-plane and (c) out-of-plane phonons. (d) Schematic structures of phonon-based field effect transistor (PFET) composed of Gr and h-BN gating.

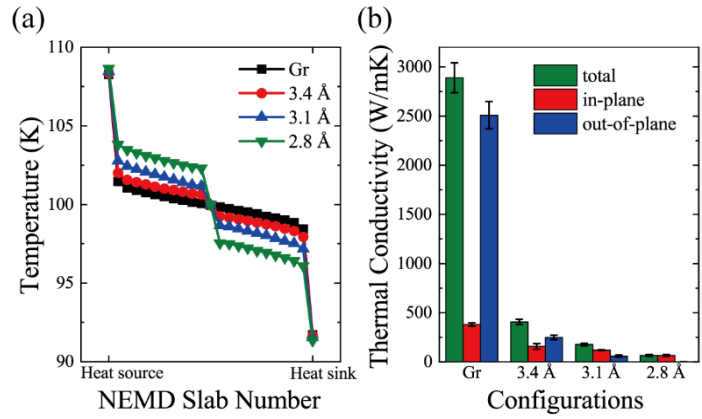

FIG. 2. (a) Temperature profile as a function of NEMD groups for freestanding Gr and PFET composed of Gr and h-BN gating with different $d_{int}$ (= 3.4 Å, 3.1 Å, 2.8 Å) at 100K. (b) Total thermal conductivity, thermal conductivity from in-plane and out-of-plane phonons for freestanding Gr and PFET composed of Gr and h-BN gating with different $d_{int}$ at 100K.

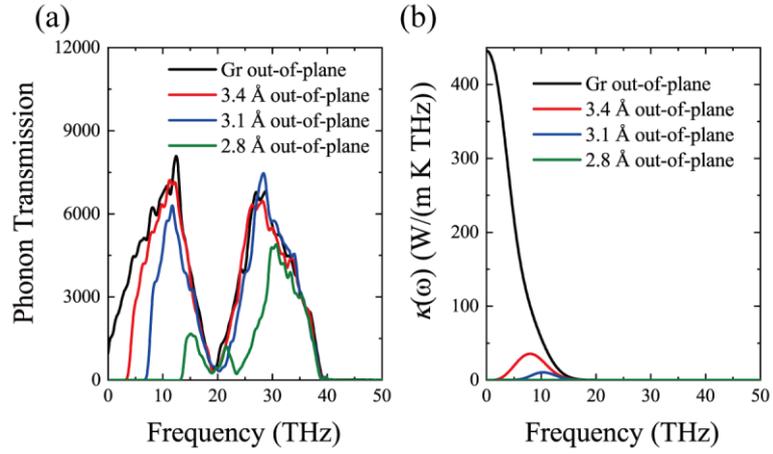

FIG. 3. (a) Spectral phonon transmission and (b) Spectral thermal conductivity of out-of-plane modes at 100K for freestanding Gr and PFET composed of Gr and h-BN gating with different $d_{int}$ (= 3.4 Å, 3.1 Å, 2.8 Å).

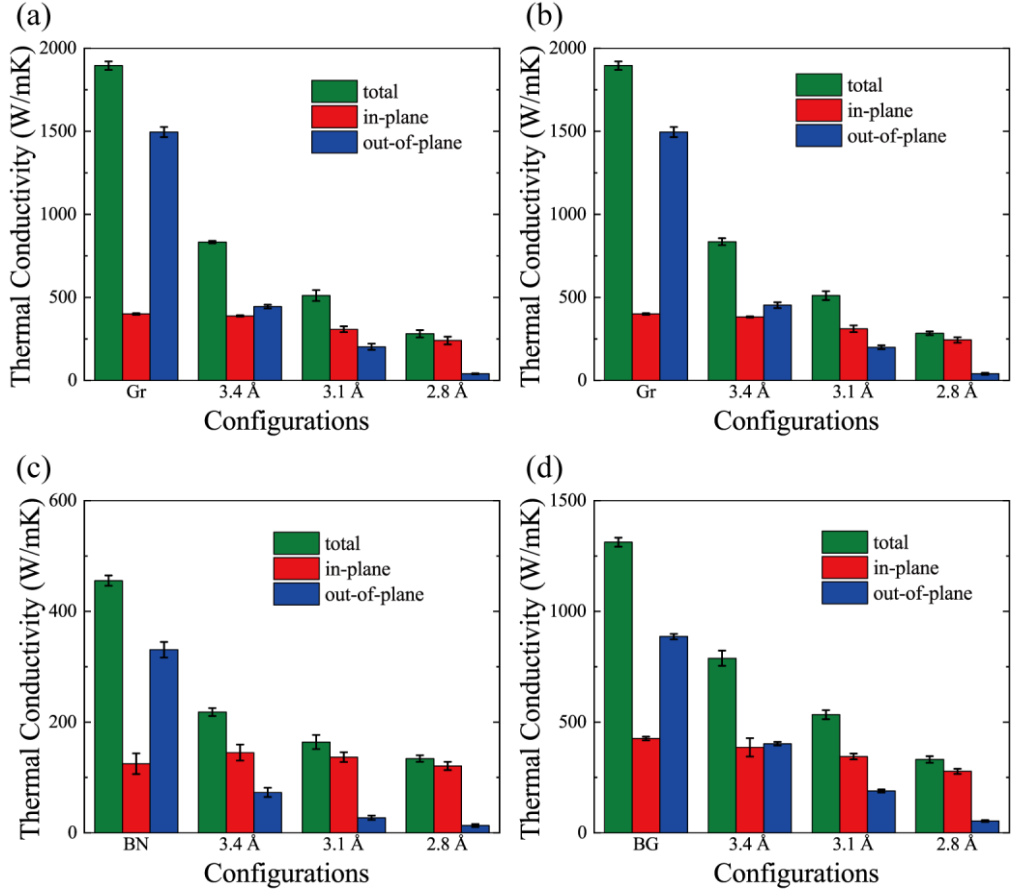

FIG. 4. (a)-(d): Total thermal conductivity, thermal conductivity from in-plane and out-of-plane phonons at 300 K for (a) Gr and PFET composed of Gr and h-BN gating with different $d_{int}$ (= 3.4 Å, 3.1 Å, 2.8 Å). (b) Gr and PFET composed of Gr and Gr gating with different $d_{int}$. (c) h-BN and PFET composed of h-BN and h-BN gating with different $d_{int}$. (d) BG and PFET composed of BG and h-BN gating with different $d_{int}$.